\documentstyle[12pt]{article}
\newcommand{\beq}{\begin{equation}}
\newcommand{\eeq}{\end{equation}}
\newcommand{\bea}{\begin{eqnarray}}
\newcommand{\eea}{\end{eqnarray}}

\begin{document}
\setcounter{page}{0}
\topmargin 0pt
\oddsidemargin 5mm
\renewcommand{\thefootnote}{\fnsymbol{footnote}}
\newpage
\setcounter{page}{0}
\begin{titlepage}

\begin{flushright}
QMW-PH-97-24\\
{\bf hep-th/9708021}\\
 {\it August 1997}
\end{flushright}
\vspace{0.5cm}
\begin{center}
{\Large {\bf On the Schild action for $D=0$ and $D=1$ strings}} \\
\vspace{1.8cm}
\vspace{0.5cm}
{ Oleg A.
Soloviev\footnote{e-mail: O.A.Soloviev@QMW.AC.UK}} \\
\vspace{0.5cm}
{\em Physics Department, Queen Mary and Westfield College, \\
Mile End Road, London E1 4NS, United Kingdom}\\
\vspace{0.5cm}
\renewcommand{\thefootnote}{\arabic{footnote}}
\setcounter{footnote}{0}
\begin{abstract}
{It is shown that the integration measure over the matrix $Y$ in the matrix representation of the 
Schild action can be fixed by comparing the Schild matrix model with the random lattice string 
model for $D=0$. It is further checked that the given measure is consistent with the case $D=1$ 
as well. 
}
\end{abstract}
\vspace{0.5cm}
 \end{center}
\end{titlepage}
\newpage
\section{Introduction}
 
Since Polyakov's seminal paper \cite{Polyakov} on Quantum geometry of bosonic strings, there has 
been a tremendous effort to solve string theory exactly. Unfortunately, it has been realised that 
Polyakov's continuous approach does not take us any further than a perturbative description. For 
this reason other alternative formulations of strings have recently attracted a great deal of 
attention. Among them, Schild's action \cite{Schild} seems to be  the most promising candidate 
for a nonperturbative description of string theory. A remarkable fact is that both Polyakov's and 
Schild's actions can be derived from one theory \cite{Yoneya}. In this sense these two models are 
dual to each other even though their field contents are different off-shell. It is argued that 
the given duality can be an exact equivalence at the quantum level \cite{Yoneya}. 

A dramatic breakthrough occured when it was realised that the Schild action of type IIB 
superstring allows a natural matrix representation in the large $N$-limit\cite{IKKT}. For early 
ideas about the relation between the Schild string and matrix models see 
\cite{Zachos},\cite{Bars} which also can be connected to a proposal on quantisation of the Schild 
action in ref. \cite{Eguchi}. However, the IKKT matrix approach to the Schild formulation of 
(super)strings suffers one subtlety concerning the limit $N\to\infty$. Namely, in the IKKT matrix 
theory at the quantum level the size $N$ of matrices is treated as a dynamical variable 
\cite{IKKT}. This seems to contradict the main idea of taking the large $N$ limit in the matrix 
model. The mentioned subtlety has been cured in a modified matrix model proposed in \cite{FMOSZ} 
where a new matrix variable $Y$ has been introduced so that the limit $N\to\infty$ can be taken 
straightforwardly. The curious fact is that $Y$ does not couple to the fermionic matrices. 
Therefore, it appears that some of its properties can be studied in a pure bosonic theory.

The new matrix description suggested in \cite{IKKT},\cite{FMOSZ} provides us with a new tool of 
investigating nonperturbative properties of string theory. The hope is that these new matrix 
models can overcome the problems of the old matrix model approach to nonperturbative quantisation 
of strings \cite{old} broadly studied a few years ago. The remarkable success of the old matrix 
models was a nonperturbative description of strings in dimensions $D\le1$. The major difficulties 
occured in an attempt to extend this approach to string theories in $D>1$ and superstrings.

If correct the new matrix models have to solve the given problems and at the same time they have 
to reproduce the exact results obtained for $D\le1$ strings within the old approach. The 
consistency between the new and the old matrix theories is required for justification of the 
equivalence between the Polyakov and the Schild formulations of string theory as suggested in 
\cite{Yoneya}. This poses a question about the relation between these two matrix model approaches 
which has already been raised in \cite{Bars}. One nontrivial result of such a relation has been 
discussed in \cite{Kristjanson}.

The aim of this paper is to compare the new matrix model with the old matrix model description of 
the bosonic string in dimensions $D=0$ and $D=1$. By doing so we clarify some issues concerning 
the modified matrix theory constructed in \cite{FMOSZ}. In particular, we shall focus on the form 
of the potential for the matrix $Y$ which cannot be fixed from the quasiclassical consideration. 
This potential defines the measure of the matrix integral over $Y$ \cite{Chekhov1}. From this 
point of view our paper is a further extension of the analysis started in \cite{Chekhov1}.

The paper is organized as follows. In section 2 we discuss the classical relation between 
Polyakov's and Schild's formulations of the Nabu-Goto string. Also we summarize the main ideas of 
introducing the matrix representation for the Schild action and implications of this 
representation for boundary conditions of the continuous theory. In section 3 we continue to 
study the quantization of the $D=0$ Schild action in the matrix representation. We show that the 
comparison with the $D=0$ Polyakov string allows us to fix the form of the potential of the 
matrix $Y$. This potential is further checked in the case of $D=1$ string. We conclude in section 
4.

\section{From strings to matrices}

There exists an infinite number of classical functionals all of which give rise to one and the 
same equations of motion of strings. However, among them two formulations are distinguished 
by their simplicity and elegance. One is due to Polyakov \cite{Polyakov} and the other due to 
Schild \cite{Schild}.

The Polyakov formulation is based on the following elegant action
\begin{equation}
S_{Polyakov}=-{1\over\kappa}\int d^2\xi\sqrt{-
g}g^{ab}\partial_aX\cdot\partial_bX,\label{polyakov}\end{equation}
where $X^\mu$, $\mu=1,2,...,D$, are coordinates of the string in the target space-time and 
$g^{ab}$ is the two-dimensional metric on the string world sheet parametrized by $\xi^1$ and 
$\xi^2$. $\kappa=4\pi\alpha'$ is a constant related to the string tension. 

The Schild action is defined as follows
\begin{equation}
S_{Schild}=-\int d^2\xi\left(-
{1\over2\kappa^2e}\;\Sigma^2~+~e\right).\label{schild}\end{equation}
Here
\begin{equation}
\Sigma^{\mu\nu}=\epsilon^{ab}\partial_aX^\mu\partial_bX^\nu.\label{sigma}\end{equation}
The variable $e$ is a positive definite scalar density defined on the world sheet.

It is easy to check that by excluding the metric $g^{ab}$ in the Polyakov action and the density 
$e$ in the Schild action one arrives at the standard Nabu-Goto action. Moreover the two models 
are dual to each other \cite{Yoneya}. Indeed, let us consider the following functional
\begin{equation}
S(t,e,X)=\int d^2\xi{1\over e}\left(\det 
t~+~{1\over\kappa}t^{ab}\partial_aX\cdot\partial_bX\right)~-~\int 
d^2\xi\;e,\label{functional}\end{equation}
where the new variable $t^{ab}$ was introduced.

The equivalence of $S(t,e,X)$ to the Schild action follows from the identity
\begin{equation}
S(t,e,X)=S_{Schild}(e,X)~+~\int d^2\xi\;{1\over e}\det\tilde t,\label{toschild}\end{equation}
where
\begin{equation}
\tilde t^{ab}=t^{ab}+{1\over\kappa}\epsilon^{ac}\epsilon^{bd}\partial_cX\cdot\partial_dX.
\label{ttilde}\end{equation}
Now it is easy to see that upon the equation of motion of $\tilde t^{ab}$, which is
\begin{equation}
\tilde t^{ab}=0,\label{eqttilde}\end{equation}
the functional $S(t,e,X)$ reduces to the Schild action (\ref{schild}) of $e$ and $X^\mu$.

The equivalence to the Polyakov action is also straightforward. It is convenient to make a change 
of variables:
\begin{eqnarray}
t^{ab}&=&e^2\;g^{ab},\nonumber\\ & & \\
e&=&\tilde e\sqrt{-g}.\nonumber\label{change}\end{eqnarray}
Then
\begin{equation}
S(t,e,X)\to S(\tilde e,g^{ab},X)=\int d^2\xi\;\tilde e(\tilde e^2-1)\sqrt{-
g}~+~{1\over\kappa}\int d^2\xi\;\tilde e\sqrt{-
g}g^{ab}\partial_aX\cdot\partial_bX.\label{topolyakov}\end{equation}
Correspondingly, the equation of motion of $g^{ab}$ is given as follows
\begin{equation}
{1\over2}\tilde e(\tilde e^2-1)g_{ab}~+~{1\over\kappa}\left(\partial_aX\cdot\partial_bX-
{1\over2}g_{ab}g^{cd}\partial_cX\cdot\partial_dX\right)=0.\label{eqg}\end{equation}
Take the trace of the given equation of motion. We get
\begin{equation}
\tilde e(\tilde e^2-1)=0,\label{tildee}\end{equation}
which implies
\begin{equation}
\tilde e=1,\label{e=1}\end{equation}
since we assumed that $e>0$.

Thus, if we plug solution (\ref{e=1}) back into formula (\ref{topolyakov}), we obtain the 
Polyakov action for $g^{ab}$ and $X^\mu$. This completes the prove of the classical equivalence 
between the Polyakov and the Schild formulations.

At the quantum level the situation is much more complex\footnote{The quantum equivalence between 
the Nabu-Goto action and the Schild action has been studied in \cite{Spallucci}.}. The formal 
manipulations under the functional integral discussed in \cite{Yoneya} are probably correct only 
for the critical strings. For noncritical strings the analysis gets stuck because of lack of a 
nonperturbative definition of the functional measure. However, as we shall see in some cases the 
comparison of the two models can be done even in noncritical dimensions. These will be the cases 
when both the theories are represented as matrix models.

Now we want to turn to the matrix description of the Schild action. It starts with an observation 
that the latter takes a pure algebraic form in terms of the following Poisson brackets 
\cite{Eguchi},\cite{Bars}
\begin{equation}
\{A,B\}\equiv{1\over e}\epsilon^{ab}\partial_aA\partial_bB.\label{bracket}\end{equation}
The next drastic step is to replace all Poisson brackets by commutators of $N\times N$ matrices 
\cite{IKKT} which is justifiable only in the limit $N\to\infty$. A subtlety arises in passing 
from the two dimensional integral over the world sheet to the matrix trace. In ref.\cite{IKKT} 
this transition is defined according to the following rule
\begin{eqnarray}
\{A,B\}&\to&[A,B],\nonumber\\ & & \\
\int d^2\xi\;e&\to&\mbox{Tr}.\label{jpass}\nonumber\end{eqnarray}

With the given rule the Schild action is replaced by the following matrix model \cite{IKKT}
\begin{equation}
S_{IKKT}=-{\alpha\over4}\mbox{Tr}[X^\mu,X^\nu]^2~+~\beta\mbox{Tr}{\bf 
1},\label{IKKT}\end{equation}
where $\alpha$ and $\beta$ are some constants of order $N$. As one can see this matrix model does 
not have a matrix variable for the density $e$ of the continuous theory. According to \cite{IKKT} 
the size $N$ of matrices has to be treated as such an additional dynamical degree of freedom. 

In what follows we shall study a modified matrix model \cite{FMOSZ} which we shall call Schild 
matrix model (SMM). Its classical action is given as follows
\begin{equation}
S_{SMM}=-{\alpha\over4}\mbox{Tr}\;Y^{-
1}[X^\mu,X^\nu]^2~+~\beta\mbox{Tr}\;Y,\label{FMOSZ}\end{equation}
where $Y$ is a matrix variable associated with the field $e$. The given matrix theory can be 
derived from the Schild action by the following rule \cite{FMOSZ}
\begin{eqnarray}\label{rule}
e&\to&Y,\nonumber\\
\{X^\mu,X^\nu\}_{P.B.}\equiv\epsilon^{ab}\partial_aX^\mu\partial_bX^\nu&\to&[X^\mu,X^\nu]\\
\int d^2\xi&\to&\mbox{Tr}.\nonumber\end{eqnarray}

This rule (\ref{rule}) contains some nontrivial information about boundary conditions of the 
world sheet. Indeed, according to the standard property of the trace one can derive the following 
relation
\begin{equation}
\mbox{Tr}[X^\mu,X^\nu]=0.\label{classical}\end{equation}
In terms of continuous strings this means
\begin{equation}
\int d^2\xi\;\{X^\mu,X^\nu\}_{P.B.}=\int 
d^2\;\epsilon^{ab}\partial_aX^\mu\partial_bX^\nu=\int_\Gamma 
ds\;X^\mu\partial_tX^\nu=0,\label{boundary}\end{equation}
where $\partial_t$ is a derivative tangential to the boundary $\Gamma$. Eq.(\ref{boundary}) is 
fulfilled when $X^\mu$ obeys the following boundary condition
\begin{equation}                                            
\partial_tX^\mu|_\Gamma=0\label{dirichlet}\end{equation}
In other words, if there are boundaries on the world sheet, all $X^\mu$'s have to satisfy the 
Dirichlet boundary conditions. This observation supports a conjecture that type IIB matrix model 
(whose bosonic part is described by the SMM) is the effective action of $N$ D-instantons 
\cite{Witten},\cite{IKKT}. It is even more interesting to look at the D-brane solutions of the 
(super)-SMM which are given as follows \cite{IKKT}
\begin{equation}
[B^\alpha,B^\beta]=-ig^{\alpha\beta}{\bf 1},\label{dbranes}\end{equation}
where $B^\alpha$, $\alpha=0,1,...,p,$ (for odd p) are world volume coordinates of a p-brane 
\cite{IKKT},\cite{FMOSZ} and $g^{\alpha\beta}$ is some antisymmetric matrix. The rest of the 
coordinates can be set to zero. Solutions of this type make sense only in the large $N$ limit. In 
terms of the Poisson brackets eq.(\ref{dbranes}) is equivalent to the following relation
\begin{equation}
\int d^2\xi\;\{B^\alpha,B^\beta\}_{P.B.}=\int_\Gamma 
ds\;B^\alpha\partial_tB^\beta\ne0.\label{ne}\end{equation}
In order for this to be the case, $B^\alpha$ have to satisfy the following boundary condition
\begin{equation}
\partial_nB^\alpha|_\Gamma=0,\label{neumann}\end{equation}
where $\partial_n$ is a derivative normal to the boundary. Indeed, it is well known that the only 
natural boundary conditions in string theory are either Neumann or Dirichlet conditions 
\cite{Alvarez}. Since $\partial_tB^\alpha\ne0$ on the boundary, condition (\ref{neumann}) has to 
be imposed.
This is nothing but the Neumann boundary condition as it should be in the case of D-branes 
\cite{Dai}. All other coordinates still have to obey the Dirichlet boundary conditions. Moreover, 
eq.(\ref{ne}) signals that the solutions of type (\ref{dbranes}) require the presence of 
boundaries on the world sheet. Thus, these solutions can be understood as a nonperturbative 
effect of the string interaction: boundaries are generated in the process of multi-string 
coupling. Multi-string states are inevitable ingredients of the given matrix description due to 
the claster decomposition of matrices $X^\mu$.

\section{{D=0} and {D=1} SMM}

The matrix representation of the Schild action allows us to define a nonperturbative quantization 
of strings in terms of large $N$ matrix integrals \cite{IKKT},\cite{FMOSZ}. This is a nontrivial 
task because the integration over the matrix $Y$ requires a properly defined measure which cannot 
be obtained in the quasiclassical approximation. The measure can be understood as a potential 
$V(Y)$ in the SMM \cite{Chekhov1}. Therefore, we write down the following general expression for 
the quantum SMM
\begin{equation}
S_{SMM}=N\left[\mbox{Tr}\;Y^{-1}\Lambda~+~V(Y)\right],\label{QSMM}\end{equation}
where
\begin{equation}
\Lambda\sim[X^\mu,X^\nu]^2.\label{Lambda}\end{equation}

The form of the potential $V(Y)$ has been discussed in \cite{FMOSZ},\cite{Chekhov1}. In 
\cite{Chekhov1} it is argued that the potential $V(Y)$ has to be fixed as follows
\begin{equation}
V_{FMOSZ-CZ}(Y)=\beta\mbox{Tr}Y~+~\gamma\mbox{Tr}\ln Y,\label{guess}\end{equation}
where it is essential to choose the parameter $\gamma$ of order one \cite{Chekhov1}:
\begin{equation}
\gamma=1~+~{\cal O}(1/N).\label{gamma}\end{equation}
The argument is based on the locality and reparametrisation invariance of the corresponding 
effective action (obtained by integrating out the matrix $Y$).

In what follows we would like to put forward more arguments in favour of a further modification 
of the potential $V(Y)$. To this end we would like to turn to two particular cases of $D=0$ and 
$D=1$ SMM's. 

Let us start with the $D=0$ SMM. Because there are no space-time coordinates $X^\mu$, the 
corresponding partition function is completely defined by the potential $V(Y)$:
\begin{equation}
Z_{SMM}(D=0)=\int{\cal D}Y\;\exp[-N\;V(Y)],\label{d=0}\end{equation}
where ${\cal D}Y$ is the standard flat measure. This partition function is supposed to give rise 
to the nonperturbative expression for free energy.

At the same time, following the random matrix representation of the $D=0$ Polyakov string 
\cite{Kostov},\cite{Brezin}, one can write down the explicit expression for the corresponding 
$D=0$ free energy in terms of the integral over a hermitian matrix
\begin{equation}
Z_P(D=0)=\int{\cal D}M\;\exp\left[-N\;\mbox{Tr}\left({1\over2}M^2~-
~gM^4\right)\right],\label{kostov}\end{equation}
where the coupling constant $g$ approaches the critical value
\begin{equation}
g\to g_c={1\over48}.\label{g}\end{equation}

In order to compare $Z_P(D=0)$ with $Z_{SMM}(D=0)$, we present the former in the following 
equivalent form \cite{Kostov}
\begin{equation}
Z_P(D=0)=const\;\int{\cal D}U\;\exp\left\{-N\;\mbox{Tr}\left[{1\over2}U^2~+~\ln(1-\lambda 
U)\right]\right\},\label{newkostov}\end{equation}
where
\begin{equation}
\lambda=\sqrt{8g}.\label{lambda}\end{equation}
The important point to be made is that the given matrix model possesses a remarkable symmetry 
known as the face/vertex duality of the Feynmann graphs \cite{Kostov},\cite{Siegel1} which is 
related to the T-duality of the continuous string theory \cite{Siegel2},\cite{Gross}.

It is convenient to further change variables
\begin{equation}
1-\lambda U\to\lambda^2Y,\label{why}\end{equation}
in terms of which the potential in eq.(\ref{newkostov}) takes the following form
\begin{equation}
V(Y)=N\;\mbox{Tr}\left[-Y~+~\ln 
Y~+~{\lambda^2\over2}Y^2\right]~+~\delta,\label{potential}\end{equation}
where
\begin{equation}
\delta=N^2\left({1\over2\lambda^2}~+~\ln\lambda^2\right).\label{delta}\end{equation}

The first two terms in eq.(\ref{potential}) are nothing but the Penner model at the critical 
point \cite{Penner},\cite{Distler}
\begin{equation}
V_{Penner}(Y)=Nt\;\mbox{Tr}\;\left(\ln\; Y~-~Y\right),\label{penner}\end{equation}
where
\begin{equation}
t\to t_c=1.\label{t}\end{equation}
This part of the potential $V(Y)$ is in agreement with the expression proposed in 
ref.\cite{FMOSZ}  and given by eq.(\ref{guess}). However, there is the Gaussian term in 
eq.(\ref{potential}) which can be understood as a correction to formula (\ref{guess}) in the 
limit $\lambda\to0$ (as advocated in \cite{Soloviev}). 

As one can see the equivalence between the Polyakov $D=0$ string and the $D=0$ SMM requires the 
modified potential given by eq.(\ref{potential}). The nature of the corrected potential can be 
further clarified in the case of $D=1$ string. 

Remarkably, the $D=1$ SMM is described like the $D=0$ case by the potential alone\footnote{It 
might be interesting to compare the given similarity between the $D=0$ and $D=1$ descriptions and 
the phase transition between the $c=0$ and $c=1$ CFT's coupled to 2D gravity discovered in 
\cite{Gross},\cite{Yang}}:
\begin{equation}
S_{SMM}(D=1)=N\;V(Y).\label{d=1}\end{equation}
Indeed, the matrix $\Lambda$ in eq.(\ref{QSMM}) vanishes for a single $X$,
\begin{equation}
\Lambda\sim[X,X]^2=0.\label{vanishing}\end{equation}

Thus the $D=1$ SMM partition function is defined according to
\begin{equation}
Z_{SMM}(D=1)=\int{\cal D}X\int{\cal D}Y\;{\rm e}^{-N\;V(Y)}.\label{D=1}\end{equation}
It differs from formula (\ref{d=0}) by the additional integral over the matrix $X$. Let us 
introduce the following notation
\begin{equation}
\nu\equiv\int{\cal D}X.\label{nu}\end{equation}
Then the partition function can be presented as follows
\begin{equation}
Z_{SMM}(D=1)=\nu{\rm e}^{-\delta}\;\int{\cal D}Y\;\exp\left[-N\;{\rm Tr}\left(
\ln Y~-~Y~+~{\lambda^2\over2}Y^2\right)\right],\label{c=1}\end{equation}
where $\delta$ is given by eq.(\ref{delta}).

Let us take the limit 
\begin{equation}
\lambda\to0.\label{limit}\end{equation}
It is easy to see that in this limit the constant $\delta$ behaves as follows
\begin{equation}
\delta\to{N^2\over2\lambda^2}.\label{limdelta}\end{equation}
Consider the following integral
\begin{equation}
Z(\eta)={\rm e}^{-{N\;{\rm Tr}\eta^2\over2}}\;\int{\cal D}Y\;\exp\left[-N\;{\rm Tr}\left(
\ln Y~-~Y~+~{1\over2}\eta^{-1}Y\eta^{-1}Y\right)\right],\label{integral}\end{equation}
where
\begin{equation}
\eta={\rm diag}(1/\lambda,1/\lambda,...,1/\lambda).\label{eta}\end{equation}
Obviously,
\begin{equation}
Z_{SMM}(D=1)=\nu\;Z(\eta).\label{obvious}\end{equation}

The beautiful thing is that $Z(\eta)$ satisfies the Schwinger-Dyson equation \cite{Chekhov2}
\begin{equation}
\left({\partial^2\over\partial\eta^2}~+~N\eta{\partial\over\partial\eta}~-
~N^2\right)\;Z(\eta)=0,\label{dyson}\end{equation}
which is equivalent to the Virasoro constraints of the generalized Kontsevich-Penner model (or 
the Gaussian Kontsevich model) \cite{Chekhov2},\cite{Chekhov3}. Correspondingly, eq.(\ref{eta}) 
has to be thought of as being a constraint on solutions of eq.(\ref{dyson}) with the arbitrary 
matrix $\eta$. Such solutions do exist \cite{Chekhov2},\cite{Chekhov3}. This model has been 
extensively studied in \cite{Chekhov2},\cite{Chekhov3},\cite{Chekhov4}, see also \cite{Morozov}. 
It has been shown that in the limit $\lambda\to0$ this theory coincides with the standard Penner 
model. The latter is known to describe the $c=1$ CFT (Polyakov's $D=1$ string) compactified on a 
circle of the self-dual radius \cite{Distler},\cite{Chaudhuri}. It is remarkable that exactly at 
the self-dual radius the quantity $\nu$ in eq.(\ref{nu}) is equal to one.

All in all we arrive at the conclusion that the Polyakov $D=1$ string and the $D=1$ SMM are 
equivalent if we choose the potential $V(Y)$ in form (\ref{potential}). Moreover, the equivalence 
requires the specific self-dual value of the radius of the compact dimension. The latter may 
imply that a $D=1$ string does not exist at different values of the compactification.

\section{Conclusion}

We have shown that a simple consideration of the SMM at $D=0$ allows us to fix the potential 
$V(Y)$ in the following form
\begin{equation}
V(Y)=N\;{\rm Tr}\left[{Y^2\over2}~+~\ln(1-\lambda Y)\right].\label{gaussian}\end{equation}
This matrix theory is known as the Gaussian Kontsevich model. It is invariant under the 
face/vertex duality which seems to play a significant role in the nonperturbative understanding 
of string theory. The given potential modifies the anzats suggested in \cite{FMOSZ} when 
$\lambda$ is not vanishing. The critical value of the matrix model coupling constant $\lambda$ 
depends on the dimensionality of $X^\mu$'s. It is very interesting to understand how the constant 
$\lambda$ is related to the string coupling constant $g_s$. One plausible conjecture is that
\begin{equation}
\lambda=\xi~+~{\cal O}(\xi^2),\label{conj}\end{equation}
where
\begin{equation}
\xi={g_s\over1+g^2_s}.\label{xi}\end{equation}
In this case, the SMM is consistent with the symmetry $g_s\to1/g_s$ which is thought of as being 
a nonperturbative symmetry of IIB superstring \cite{Hull},\cite{Schwarz}.

In dimensions $D>1$ the SMM is formulated as follows
\begin{equation}
S_{SMM}=N\;{\rm Tr}\left[(1-\lambda Y)^{-1}\Lambda~+~{Y^2\over2}~+~\ln(1-\lambda 
Y)\right].\label{D}\end{equation}
This matrix model deserves further investigation.

I thank PPARC for financial support.


\begin{thebibliography}{99}
\bibitem{Polyakov} A. M. Polyakov, Phys. Lett. {\bf B103} (1981) 207.
\bibitem{Schild} A. Schild, Phys. Rev. {\bf D16} (1977) 1722.
\bibitem{Yoneya} T. Yoneya, {\it Schild action and space-time uncertainty principle in string 
theory}, hep-th/9703078.
\bibitem{IKKT} N. Ishibashi, H. Kawai, Y. Kitazawa and A. Tsuchiya, {\it A large-N reduced model 
as superstring}, KEK-TH-503, TIT/HEP-357, hep-th/9612115.
\bibitem{Zachos} D. B. Fairlie, P. Fletcher and C. K. Zachos, J. Math. Phys. {\bf 31} (1990) 
1088; Phys. Lett. {\bf B405} (1997) 37, hep-th/9704037.
\bibitem{Bars} I. Bars, Phys. Lett. {\bf B245} (1990) 35, see also hep-th/9706177.
\bibitem{Eguchi} T. Eguchi, Phys. Rev. Lett. {\bf 44} (1980) 126.
\bibitem{FMOSZ} A. Fayyazuddin, Y. Makeenko, P. Olesen, D. J. Smith and K. Zarembo, {\it Towards 
a non-perturbative formulation of IIB superstrings by matrix models}, NBI-HE-97-09, ITEP-TH-
09/97, hep-th/9703038, to be published in Nucl. Phys. {\bf B}.
\bibitem{old} E. Br\'ezin and V. A. Kazakov, Phys. Lett. {\bf B236} (1990) 144;
M. Douglas and S. Shenker, Nucl. Phys. {\bf B335} (1990) 635;
D. J. Gross and A. A. Migdal, Phys. Rev. Lett. {\bf 64} (1990) 127.
\bibitem{Kristjanson} C. F. Kristjanson and P. Olesen, {\it A possible IIB superstring matrix 
model with Euler characteristic and a double scaling limit}, hep-th/9704017.
\bibitem{Witten} E. Witten, Nucl. Phys. {\bf B460} (1996) 335, hep-th/9510135.
\bibitem{Chekhov1} L. Chekhov and K. Zarembo, {\it Effective action and measure in matrix model 
of type IIB superstrings}, hep-th/9705014.
\bibitem{Spallucci} S. Ansoldi, A. Aurilia and E. Spallucci, Phys. Rev. {\bf D53} (1996) 870, 
hep-th/9510133; {\it Hausdorff dimension of a quantum string}, hep-th/9705010, to appear in Phys. 
Rev. {\bf D}.
\bibitem{Alvarez} O. Alvarez, Nucl. Phys. {\bf B216} (1983) 125.
\bibitem{Dai} J. Dai, R. G. Leigh and J. Polchinski, Mod. Phys. Lett. {\bf A4} (1989) 2073.
\bibitem{Kostov} I. K. Kostov and M. L. Mehta, Phys. Lett. {\bf B189} (1987) 118.
\bibitem{Brezin} E. Br\'ezin and V. A. Kazakov, Phys. Lett. {\bf B236} (1990) 144.
\bibitem{Siegel1} W. Siegel, Phys. Rev. {\bf D54} (1996) 2797, hep-th/9603030.
\bibitem{Siegel2} W. Siegel, Phys. Lett. {\bf B252} (1990) 558.
\bibitem{Gross} D. J. Gross and I. Klebanov, Nucl. Phys. {\bf B344} (1990) 475.
\bibitem{Penner} R. C. Penner, J. Diff. Geom. {\bf 27} (1988) 35.
\bibitem{Distler} J. Distler and C. Vafa, Mod. Phys. Lett. {\bf A6} (1991) 259.
\bibitem{Soloviev} O. Soloviev, {\it Random lattice strings vs. type IIB matrix models}, hep-
th/9707043.
\bibitem{Yang} Z. Yang, Phys. Lett. {\bf B243} (1990) 365.
\bibitem{Chekhov2} L. Chekhov and Yu. Makeenko, Mod. Phys. Lett. {\bf A7} (1992) 1223, hep-
th/9201033.
\bibitem{Chekhov3} L. Chekhov and Yu. Makeenko, Phys. Lett. {\bf B278} (1992) 271, hep-
th/9202006.
\bibitem{Chekhov4} L. Chekhov, {\it Matrix models and geometry of moduli spaces}, hep-th/9509001.
\bibitem{Morozov} A. Morozov, {\it Matrix models as integrable systems}, hep-th/9502091.
\bibitem{Chaudhuri} S. Chaudhuri, H. Dykstra and J. Lykken, Mod. Phys. Lett. {\bf A6} (1991) 
1665.
\bibitem{Hull} C. M. Hull, Phys. Lett. {\bf B357} (1995) 545.
\bibitem{Schwarz} J. H. Schwarz, Phys. Lett. {\bf B360} (1995) 13; Erratum: ibid. {\bf B364} 
(1995) 252, hep-th/9508143.
\end{thebibliography}
\end{document}